\documentclass[a4paper,11pt]{article}
\usepackage[latin1]{inputenc}
\usepackage{amssymb}

\title{{\bf  Three Applications to Rational Relations of the High Undecidability of the Infinite Post Correspondence Problem in a Regular $\om$-Language}}

\author{Olivier Finkel   \\{\it   Equipe de Logique Math\'ematique}
  \\ CNRS et  Universit\'e Paris Diderot Paris 7
 \\ UFR de Math\'ematiques case 7012, site Chevaleret,\\75205 Paris Cedex 13, 
 France.\\ finkel@logique.jussieu.fr 
}

\date{}
\begin{document}

\newtheorem{The}{Theorem}[section]
\newtheorem{Pro}[The]{Proposition}
\newtheorem{Deff}[The]{Definition}
\newtheorem{Lem}[The]{Lemma}
\newtheorem{Rem}[The]{Remark}
\newtheorem{Exa}[The]{Example}
\newtheorem{Cor}[The]{Corollary}
\newtheorem{Not}[The]{Notation}

\newcommand{\fa}{\forall}
\newcommand{\Ga}{\mathsf\Gamma}
\newcommand{\Gas}{\Ga^\star}
\newcommand{\Gao}{\Ga^\omega}

\newcommand{\Si}{\mathsf\Sigma}
\newcommand{\Sis}{\Si^\star}
\newcommand{\Sio}{\Si^\omega}
\newcommand{\ra}{\rightarrow}
\newcommand{\hs}{\hspace{12mm}

\noi}
\newcommand{\lra}{\leftrightarrow}
\newcommand{\la}{language}
\newcommand{\ite}{\item}
\newcommand{\Lp}{L(\varphi)}
\newcommand{\abs}{\{a, b\}^\star}
\newcommand{\abcs}{\{a, b, c \}^\star}
\newcommand{\ol}{ $\omega$-language}
\newcommand{\orl}{ $\omega$-regular language}
\newcommand{\om}{\omega}
\newcommand{\nl}{\newline}
\newcommand{\noi}{\noindent}
\newcommand{\tla}{\twoheadleftarrow}
\newcommand{\de}{deterministic }
\newcommand{\proo}{\noi {\bf Proof.} }
\newcommand {\ep}{\hfill $\square$}
\renewcommand{\thefootnote}{\star{footnote}} 
\newcommand{\trans}[1]{\stackrel{#1}{\rightarrow}}
\newcommand{\dom}{\operatorname{Dom}}
\newcommand{\im}{\operatorname{Im}}

\maketitle 

\begin{abstract}
\noi  It was noticed by Harel in \cite{Harel86} that ``one can define $\Sigma_1^1$-complete versions of the well-known 
Post Correspondence Problem". 
We first give  a complete proof of this result, showing that 
 the infinite Post Correspondence Problem in a regular $\om$-language is $\Sigma_1^1$-complete, 
hence located beyond the arithmetical hierarchy and highly undecidable. 
We infer from this result   that it is $\Pi_1^1$-complete to determine whether two given infinitary rational relations are disjoint.  
Then  we prove  that there is an amazing gap between two decision problems about $\om$-rational functions realized by 
finite state B\"uchi transducers.  Indeed Prieur proved in \cite{Prieur01, Prieur02} that 
 it is decidable whether a given $\om$-rational function is continuous, while we show here that it is 
 $\Sigma_1^1$-complete to determine whether a given $\om$-rational function has at least one point of continuity.  
Next we prove that it is $\Pi_1^1$-complete to determine whether the continuity set of 
a given $\om$-rational function is $\om$-regular. This  gives the exact complexity of two problems
which were shown to be undecidable in \cite{CFS08}. 
\end{abstract}

\noi {\bf Keywords.} Decision problems; infinite Post Correspondence Problem;  analytical hierarchy;  high undecidability; 
infinitary rational relations; omega rational functions;  topology; points of continuity.

\section{Introduction}

Many classical decision problems arise  naturally in the fields of  Formal Language Theory and of Automata Theory. 
It is well known that most problems about regular languages accepted by finite automata are decidable. 
On the other hand, at the second level of the Chomsky Hierarchy, most problems about context-free languages accepted by pushdown automata or generated by 
context-free grammars are undecidable. For instance it follows from the undecidability of the Post Correspondence Problem 
that  the universality problem, the inclusion and the equivalence problems for context-free languages are also undecidable. 
Notice that some few problems about context-free languages remain decidable like the following ones: ``Is a given  context-free language $L$  empty ?" 
``Is a given  context-free language $L$ infinite ?" ``Does a given word $x$ belong to a given  context-free language $L$ ?" 
Sénizergues proved  in \cite{Senizergues01}  that  the difficult problem of the equivalence of two deterministic pushdown automata is 
decidable. 
Notice that almost proofs of undecidability results about context-free languages  rely 
on the undecidability of the Post Correspondence Problem which is complete for the class of recursively 
enumerable problems, i.e. complete at the first level of the arithmetical hierarchy. Thus undecidability proofs about context-free languages provided 
only hardness results for the first level of the arithmetical hierarchy.

\hs On the other hand, some decision problems are known to be located beyond the arithmetical hierarchy, in some classes of the analytical hierarchy, and are then 
usually called ``highly undecidable". 
Harel proved in \cite{Harel86} that many domino or tiling problems are $\Sigma_1^1$-complete or $\Pi_1^1$-complete. 
For instance the ``recurring domino problem" is $\Sigma_1^1$-complete.  It is also $\Sigma_1^1$-complete to determine whether a given Turing machine, 
when started on a blank tape, admits an infinite 
computation that reenters infinitely often in the initial state. Alur and Dill used this latter result in \cite{AlurDill94} 
to prove that the universality problem fot timed B\"uchi automata
is $\Pi_1^1$-hard. 
 In \cite{cc}, Castro and Cucker studied many decision problems for $\om$-languages of Turing machines. 
In particular, they proved that the non-emptiness and the infiniteness problems for $\om$-languages of Turing machines are $\Sigma_1^1$-complete, and that 
the universality problem, the inclusion problem, and the equivalence problem 
are $\Pi_2^1$-complete. Thus these problems are located at the first or the second level
of the analytical hierarchy.   
Using Castro and Cucker's results, some reductions of \cite{Fin-mscs06,Fin06b},  and topological arguments, 
we have proved in \cite{Fin-HI} that many decision problems 
 about   $1$-counter  $\om$-languages, context free  $\om$-languages,  or infinitary rational relations,  like  
the  universality 
problem, the inclusion problem, the equivalence problem, the determinizability problem, the complementability problem, and the 
unambiguity problem are  $\Pi_2^1$-complete.   
Notice that the exact complexity of numerous problems remains still unknown. For instance the exact complexities of the universality problem, the 
determinizability, or the complementability problem for timed B\"uchi automata which are known to be $\Pi_1^1$-hard, see \cite{AlurDill94,Fin06}.

\hs We intend to introduce here a new method for proving high undecidability results which seems to be unexplored. 
It was actually noticed by Harel in \cite{Harel86} that ``one can define $\Sigma_1^1$-complete versions of the well-known 
Post Correspondence Problem", but it seems that this possibility has not been later investigated. 
We first give  a complete proof of this result, showing that 
 the infinite Post Correspondence Problem in a regular $\om$-language is $\Sigma_1^1$-complete, 
hence located beyond the arithmetical hierarchy and highly undecidable. 
We  infer from this result  a new high undecidability result, 
proving  that it is $\Pi_1^1$-complete to determine whether two given infinitary rational relations are disjoint.  
Then we apply this  $\Sigma_1^1$-complete version of  the 
Post Correspondence Problem to the study of continuity problems for $\om$-rational functions  realized by 
finite state B\"uchi transducers,  considered  by Prieur in \cite{Prieur01, Prieur02} and by Carton, Finkel and Simonnet in \cite{CFS08}. 
We prove  that there is an amazing gap between two decision problems about $\om$-rational functions.  Indeed Prieur proved in \cite{Prieur01, Prieur02} that 
 it is decidable whether a given $\om$-rational function is continuous, while we show here that it is 
 $\Sigma_1^1$-complete to determine whether a given $\om$-rational function has at least one point of continuity.  
Next we prove that it is $\Pi_1^1$-complete to determine whether the continuity set of 
a given $\om$-rational function is $\om$-regular. This  gives the exact complexity of two problems
which were  shown to be undecidable in \cite{CFS08}. 

\hs The paper is organized as follows. We recall basic notions on automata and on the analytical hierarchy in Section 2.  We state in Section 3 the 
$\Sigma_1^1$-completeness of the infinite Post Correspondence Problem in a regular $\om$-language. We prove our main new results in Section 4. Some 
concluding remarks are given in Section 5.

\section{Recall of basic notions}

\hs We assume now  the reader to be familiar with the theory of formal ($\om$)-languages  
\cite{Thomas90,Staiger97}.
We recall some  usual notations of formal language theory. 
\nl  When $\Si$ is a finite alphabet, a {\it non-empty finite word} over $\Si$ is any 
sequence $x=a_1\ldots a_k$, where $a_i\in\Si$ 
for $i=1,\ldots ,k$ , and  $k$ is an integer $\geq 1$.  
 $\Sis$  is the {\it set of finite words} (including the empty word) over $\Si$.
 \nl  The {\it first infinite ordinal} is $\om$.
 An $\om$-{\it word} over $\Si$ is an $\om$ -sequence $a_1 \ldots a_n \ldots$, where for all 
integers $ i\geq 1$, ~
$a_i \in\Si$.  When $\sigma$ is an $\om$-word over $\Si$, we write
 $\sigma =\sigma(1)\sigma(2)\ldots \sigma(n) \ldots $,  where for all $i$,~ $\sigma(i)\in \Si$, and 
$\sigma[n] =\sigma(1)\sigma(2)\ldots \sigma(n)$. 
\nl 
 The usual concatenation product of two finite words $u$ and $v$ is 
denoted $u\cdot v$ and sometimes just $uv$. This product is extended to the product of a 
finite word $u$ and an $\om$-word $v$: the infinite word $u\cdot v$ is then the $\om$-word such that:
~~  $(u\cdot v)(k)=u(k)$  if $k\leq |u|$ , and 
 $(u\cdot v)(k)=v(k-|u|)$  if $k>|u|$.
\nl   
 The {\it set of } $\om$-{\it words} over  the alphabet $\Si$ is denoted by $\Si^\om$.
An  $\om$-{\it language} over an alphabet $\Si$ is a subset of  $\Si^\om$.  

\begin{Deff} A  B\"uchi automaton  is a 5-tuple $\mathcal{A}=(K,\Si,\delta, q_0, F)$, 
where $K$ is a finite set of states, $\Si$ is a finite input alphabet, $q_0 \in K$ is the initial state
and $\delta$ is a mapping from $K \times   \Si$ into $2^K$.  The  B\"uchi automaton  $\mathcal{A}$  is said to be  deterministic iff:
$\delta : K \times  \Si \ra K$.
\nl
Let $\sigma =a_1a_2\ldots a_n\ldots$ be an  $\om$-word over $\Si$.
A sequence of states $r=q_1q_2\ldots q_n\ldots$  is called an (infinite) run of $\mathcal{A}$ on $\sigma$, 
starting in state $p$, iff:
1) $q_1=p$  and 2) for each $i\geq 1$, $q_{i+1} \in \delta( q_i,a_i)$.
In case a run $r$ of $\mathcal{A}$ on $\sigma$ starts in state $q_0$, we call it simply ``a run of $\mathcal{A}$ 
on $\sigma$ ".
For every  run $r=q_1q_2\ldots q_n\ldots $ of $\mathcal{A}$, $In(r)$ is the set of
states in $K$ entered by $\mathcal{A}$ infinitely many times during run $r$. 
 The $\om$-language  accepted by $\mathcal{A}$ is:
$$L(\mathcal{A})= \{  \sigma\in\Si^\om  \mid \mbox{ there exists a  run } r
\mbox{ of }  \mathcal{A}  \mbox{ on } \sigma  \mbox{ such that } In(r) \cap F \neq\emptyset \}.$$

\hs An $\om$-language $L \subseteq \Si^\om$ is said to be regular iff it is accpted by some B\"uchi automaton $\mathcal{A}$. 
\end{Deff}

\noi We recall that the class of    regular   $\om$-languages  is the $\omega$-Kleene closure 
of the class of regular finitary languages, see \cite{Thomas90,Staiger97}: an $\om$-language  $L \subseteq \Si^\om$ is regular iff it is of the form 
$L=\bigcup_{1\leq i \leq n} U_i\cdot V_i^\om$, for some regular finitary languages $U_i, V_i \subseteq \Sis$. 
Acceptance of infinite words by other finite machines like pushdown automata, Turing machines, Petri nets, \ldots, with various acceptance conditions, 
has also been considered. In particular, the class of context-free $\om$-languages is the class of  $\om$-languages accepted by B\"uchi pushdown automata, 
see \cite{Thomas90,Staiger97, eh}. 

 \hs   The set of natural numbers is denoted by $\mathbb{N}$, and the set of functions from $\mathbb{N}$ into $\mathbb{N}$ is denoted by $\mathcal{F}$. 
We assume the reader to be familiar with the arithmetical   hierarchy on subsets of  $\mathbb{N}$. 
We now recall   the definition 
 of  classes of the analytical hierarchy 
which may be found in \cite{rog,Odifreddi1,Odifreddi2}. 

\begin{Deff}
Let  $k, l >0$ be some  integers. $\Phi$ is a partial computable functional of $k$ function variables and $l$ number variables if there exists $z\in \mathbb{N}$ 
such that for any $(f_1, \ldots , f_k, x_1, \ldots , x_l) \in \mathcal{F}^k \times \mathbb{N}^l$, we have 
$$\Phi (f_1, \ldots , f_k, x_1, \ldots , x_l) = \tau_z^{f_1, \ldots , f_k}(x_1, \ldots , x_l),$$
\noi where the right hand side is the output of the Turing machine with index $z$ and oracles $f_1, \ldots , f_k$ over the input $(x_1, \ldots , x_l)$. 
For $k>0$ and $l=0$, $\Phi$ is a partial computable functional  if, for some $z$, 
$$\Phi (f_1, \ldots , f_k) = \tau_z^{f_1, \ldots , f_k}(0).$$
\noi The value $z$ is called the G\"odel number or index for $\Phi$. 
\end{Deff}

\begin{Deff}
Let $k, l >0$ be some  integers and $R  \subseteq \mathcal{F}^k \times \mathbb{N}^l$. The relation $R$ is said to be a computable relation of $k$ 
function variables and $l$ number variables if its characteristic function is computable. 
\end{Deff}

\noi We now define analytical subsets of $\mathbb{N}^l$.

\begin{Deff}
A subset $R$ of $\mathbb{N}^l$ is analytical if it is computable or if there exists a computable set $S  \subseteq \mathcal{F}^m \times \mathbb{N}^n$, with 
$m\geq 0$ and $n\geq l$, such that 
$$R = \{ (x_1, \ldots , x_l) \mid (Q_1s_1)(Q_2s_2) \ldots (Q_{m+n-l}s_{m+n-l}) S(f_1, \ldots , f_m, x_1, \ldots , x_n) \}, $$
\noi where $Q_i$ is either $\fa$ or $\exists$ for $1 \leq i \leq m+n-l$, and where $s_1, \ldots , s_{m+n-l}$ are $f_1, \ldots , f_m, x_{l+1}, \ldots , x_n$ in 
some order. 
\nl The expression $(Q_1s_1)(Q_2s_2) \ldots (Q_{m+n-l}s_{m+n-l}) S(f_1, \ldots , f_m, x_1, \ldots , x_n)$ is called a predicate form for $R$. A
quantifier applying over a function variable is of type $1$, otherwise it is of type $0$. 
In a predicate form the (possibly empty) sequence of quantifiers, indexed by their type, is called the prefix of the form. The reduced prefix is the sequence of 
quantifiers obtained by suppressing the quantifiers of type $0$ from the prefix. 
\end{Deff}

\noi We can now distinguish the levels of the analytical hierarchy by considering the number of alternations in the reduced prefix. 

\begin{Deff}
For $n>0$, a $\Sigma^1_n$-prefix is one whose reduced prefix begins with $\exists^1$ and has $n-1$ alternations of quantifiers. 
A $\Sigma^1_0$-prefix is one whose reduced prefix is empty. 
For $n>0$, a $\Pi^1_n$-prefix is one whose reduced prefix begins with $\fa^1$ and has $n-1$ alternations of quantifiers. 
A $\Pi^1_0$-prefix is one whose reduced prefix is empty. 
\nl A predicate form is a $\Sigma^1_n$ ($\Pi^1_n$)-form if it has a  $\Sigma^1_n$ ($\Pi^1_n$)-prefix. 
The class of sets in some $\mathbb{N}^l$ which can be expressed in $\Sigma^1_n$-form (respectively, $\Pi^1_n$-form) is denoted by 
$\Sigma^1_n$   (respectively, $\Pi^1_n$). 
\nl The class $\Sigma^1_0 = \Pi^1_0$ is the class of arithmetical sets. 
\end{Deff}

\noi We now recall some well known results about the analytical hierarchy. 

\begin{Pro}
Let $R \subseteq \mathbb{N}^l$ for some integer $l$. Then $R$ is an analytical set iff there is some integer $n\geq 0$ such that 
$R \in \Sigma^1_n$ or $R \in \Pi^1_n$. 
\end{Pro}

\begin{The} For each integer $n\geq 1$, 
\noi 
\begin{enumerate}
\ite[(a)] $\Sigma^1_n\cup \Pi^1_n \subsetneq  \Sigma^1_{n +1}\cap \Pi^1_{n +1}$.
\ite[(b)] A set $R \subseteq \mathbb{N}^l$ is in the class $\Sigma^1_n$ iff its 
complement is in the class $\Pi^1_n$. 
\ite[(c)] $\Sigma^1_n - \Pi^1_n \neq \emptyset$ and $\Pi^1_n - \Sigma^1_n \neq \emptyset$.
\end{enumerate}
\end{The}

\noi  Transformations of prefixes  are often used, following the rules given by the next theorem. 

\begin{The}
For any predicate form with the given prefix, an equivalent predicate form with the new one can be obtained, following the 
allowed prefix transformations given below :
\noi 
\begin{enumerate}
\ite[(a)]  $\ldots \exists^0 \exists^0 \ldots \ra \ldots  \exists^0 \ldots, $ ~~~~~~~~~~~~~~ \nl $ \ldots \fa^0  \fa^0 \ldots  \ra \ldots  \fa^0 \ldots ; $
\ite[(b)]  $\ldots \exists^1 \exists^1 \ldots \ra \ldots  \exists^1 \ldots, $~~~~~~~~~~~~~~\nl  $ \ldots \fa^1  \fa^1 \ldots  \ra \ldots  \fa^1 \ldots ;$
\ite[(c)]  $\ldots \exists^0 ~~~\ldots  \ra \ldots  \exists^1 \ldots, $~~~~~~~~~~~~ \nl $ \ldots  \fa^0 ~~~\ldots  \ra \ldots   \fa^1 \ldots ; $
\ite[(d)]  $\ldots \exists^0 \fa^1 \ldots \ra \ldots \fa^1 \exists^0 \ldots$, ~~~~~~  \nl $\ldots \fa^0 \exists^1 \ldots  \ra \ldots \exists^1 \fa^0 \ldots ; $
\end{enumerate}
\end{The}

\noi We  now recall the notions of 1-reduction and of    $\Sigma^1_n$-completeness (respectively,           $\Pi^1_n$-completeness). 
Given two sets $A,B \subseteq \mathbb{N}$ we say $A$ is 1-reducible to $B$ and write $A \leq_1 B$
if there exists a total computable injective  function $f$ from      $\mathbb{N}$     to   $\mathbb{N}$        with $A = f ^{-1}[B]$. 
A set $A \subseteq \mathbb{N}$ is said to be $\Sigma^1_n$-complete   (respectively,   $\Pi^1_n$-complete)  iff $A$ is a  $\Sigma^1_n$-set 
 (respectively,   $\Pi^1_n$-set) and for each $\Sigma^1_n$-set  (respectively,   $\Pi^1_n$-set) $B \subseteq \mathbb{N}$ it holds that 
$B \leq_1 A$. 

 \hs We now recall  an example of  a  $\Sigma^1_1$-complete 
decision  problem  which will be useful in the sequel.

\begin{Deff}
A non deterministic Turing machine $\mathcal{M}$ is a $5$-tuple $\mathcal{M}=(Q, \Si, \Ga, \delta, q_0)$, where $Q$ is a finite set of states, 
$\Si$ is a finite input alphabet, $\Ga$ is a finite tape alphabet satisfying $\Si  \subseteq \Ga$ and containing a special blank symbol $\Box \in \Ga \setminus \Si$, 
$q_0$ is the initial state, 
and $\delta$ is a mapping from $Q \times \Ga$ to subsets of $Q \times \Ga \times \{L, R, S\}$. 
\end{Deff}

\noi Harel proved  the following result in  \cite{Harel86}.

\begin{The}\label{complete}
The following problem is  $\Sigma^1_1$-complete: Given a Turing machine $\mathcal{M}_z$, of index $z\in \mathbb{N}$, 
does $\mathcal{M}_z$, when started on a blank tape, admit an infinite 
computation that reenters infinitely often in the initial state $q_0$ ? 
\end{The}

\section{The infinite Post Correspondence Problem}

Recall first the well known result about the undecidability of  the Post Correspondence Problem, denoted PCP. 

\begin{The}[Post, see \cite{HopcroftMotwaniUllman2001}] Let $\Ga $  be an alphabet having at least two elements. 
Then it is undecidable
 to determine, for  arbitrary n-tuples $(x_1, x_2 \ldots ,x_n)$ and $(y_1, y_2\ldots ,y_n)$ of
 non-empty words in $\Gas$, whether there exists a non-empty sequence of indices
 $i_1, i_2 \ldots ,i_k $ such that 
$x_{i_1}x_{i_2}\ldots x_{i_k} =y_{i_1}y_{i_2}\ldots y_{i_k}$.
\end{The}

\noi On the other hand, the infinite Post Correspondence Problem, also called  $\om$-PCP,  has been shown  to be 
undecidable by Ruohonen in \cite{Ruohonen85} and by  Gire in \cite{Gire86}. 

\begin{The} Let $\Ga $  be an alphabet having at least two elements. Then it is undecidable
 to determine, for  arbitrary n-tuples $(x_1,\ldots ,x_n)$ and $(y_1,\ldots ,y_n)$ of
 non-empty words in $\Gas$, whether there exists an infinite sequence of indices
 $i_1, i_2,\ldots ,i_k \ldots $ such that 
$x_{i_1}x_{i_2}\ldots x_{i_k}\ldots  =y_{i_1}y_{i_2}\ldots y_{i_k}\ldots $ 
\end{The}

\noi Notice that an instance of the $\om$-PCP is given by two n-tuples $(x_1,\ldots ,x_n)$ and $(y_1,\ldots ,y_n)$ of
 non-empty words in $\Gas$, and  if there exist some solutions, these ones are infinite words over the alphabet $\{1, \ldots, n\}$.  

\hs 
We are going to consider now a variant of the infinite Post Correspondence Problem where we restrict solutions to $\om$-words belonging to a given 
$\om$-regular language $L(\mathcal{A})$ accepted by a given B\"uchi automaton $\mathcal{A}$. 

\hs An instance of the $\om$-PCP in a regular $\om$-language, also denoted $\om$-PCP(Reg),  is given by two n-tuples $(x_1,\ldots ,x_n)$ and $(y_1,\ldots ,y_n)$ of
 non-empty words in $\Gas$ along with a B\"uchi automaton $\mathcal{A}$ accepting words over  $\{1, \ldots, n\}$.  
 A solution of this problem  is 
an infinite sequence of indices
 $i_1, i_2,\ldots ,i_k \ldots $  such that $i_1i_2 \ldots i_k \ldots \in L(\mathcal{A})$ and 
$x_{i_1}x_{i_2}\ldots x_{i_k}\ldots  =y_{i_1}y_{i_2}\ldots y_{i_k}\ldots $. 

\hs Notice that one can 
associate in a recursive and injective way an unique integer $z$ to each B\"uchi automaton $\mathcal{A}$, this integer being 
called the index of  the automaton $\mathcal{A}$. We denote also   $\mathcal{A}_z$ the B\"uchi automaton of index $z$. 
Then each instance $I=((x_1,\ldots ,x_n), (y_1,\ldots ,y_n), \mathcal{A}_z)$ can be also characterized by an index $\bar{I}\in \mathbb{N}$.

\hs We can now state precisely  the following result. 

\begin{The}
It is $\Sigma_1^1$-complete to determine, for a given instance $I=((x_1,\ldots ,x_n), (y_1,\ldots ,y_n), \mathcal{A}_z)$, given by its index  $\bar{I}$, 
 whether there  is 
an infinite sequence of indices
 $i_1, i_2,\ldots ,i_k \ldots $  such that $i_1i_2 \ldots i_k \ldots \in L(\mathcal{A}_z)$ and 
$x_{i_1}x_{i_2}\ldots x_{i_k}\ldots  =y_{i_1}y_{i_2}\ldots y_{i_k}\ldots $. 
\end{The}

\proo 
We first prove that this problem is in the class  $\Sigma_1^1$.  It is easy to see that there is an injective computable function 
$\Phi: \mathbb{N} \ra \mathbb{N}$ such that for all $\bar{I} \in \mathbb{N}$ the B\"uchi Turing machine $\mathcal{M}_{\Phi(\bar{I})}$ of index 
$\Phi(\bar{I})$, where  $I=((x_1,\ldots ,x_n), (y_1,\ldots ,y_n), \mathcal{A}_z)$,  
accepts the set of infinite words     $i_1i_2 \ldots i_k \ldots \in L(\mathcal{A}_z)$  such that 
$x_{i_1}x_{i_2}\ldots x_{i_k}\ldots  =y_{i_1}y_{i_2}\ldots y_{i_k}\ldots $. 
Then the $\om$-PCP(Reg) of instance $I=((x_1,\ldots ,x_n), (y_1,\ldots ,y_n), \mathcal{A}_z)$ has a solution if and only if the $\om$-language of the 
B\"uchi Turing machine $\mathcal{M}_{\Phi(\bar{I})}$ is non-empty. Thus the $\om$-PCP(Reg)  is reduced to the non-emptiness problem of  
 B\"uchi Turing machines which is known to be in the class $\Sigma_1^1$. Indeed  for a given  B\"uchi Turing machine $\mathcal{M}_z$ reading infinite 
words over an alphabet $\Si$ we can express $L(\mathcal{M}_z)\neq \emptyset$ by the formula ``$\exists x \in \Sio ~ \exists r ~[ r $ is an accepting run of 
$\mathcal{M}_z$ on $x$ ]"; this is a $\Sigma_1^1$-formula because the existential second order quantifications  are followed 
by an arithmetical formula, see \cite{cc, Fin-HI} for related results. 
Therefore the  $\om$-PCP(Reg)  is also in the class $\Sigma_1^1$. 

\hs We now prove the completeness part of the theorem. Recall that the following   problem $(P)$ is $\Sigma_1^1$-complete by Theorem \ref{complete}. 

\hs $(P)$: Given a Turing machine $\mathcal{M}_z$, of index $z\in \mathbb{N}$, 
does $\mathcal{M}_z$, when started on a blank tape, admit an infinite 
computation that reenters infinitely often in the initial state $q_0$ ? 

\hs We can reduce this problem to the $\om$-PCP in a regular $\om$-language in the following way. 

\hs Let $\mathcal{M}=(Q, \Si, \Ga, \delta, q_0)$ be a Turing 
machine, where $Q$ is  a finite set of states, 
$\Si$ is a finite input alphabet, $\Ga$ is a finite tape alphabet satisfying $\Si  \subseteq \Ga$ and containing a special blank symbol $\Box \in \Ga \setminus \Si$, 
$q_0$ is the initial state, 
and $\delta$ is a mapping from $Q \times \Ga$ to subsets of $Q \times \Ga \times \{L, R, S\}$. 

\hs We are going to  associate to this Turing machine an instance of  the $\om$-PCP(Reg).  First we define the two following lists $x=(x_i)_{1\leq i \leq n}$ and 
$y=(y_i)_{1\leq i \leq n}$ of finite words 
over the alphabet $\Si \cup \Ga \cup Q \cup \{\#\}$, where \# is a symbol not in $\Si \cup \Ga \cup Q$.  

\hs 

\begin{tabular}{l|l}

x & y \\ \hline 
$\# = x_1$ & $\# q_0 \# = y_1$ \\
\# & \# \\
a & a ~~~~ ~~~~~~~~  for each $a\in \Ga$ \\ 
qa & q'b        ~~~~ ~~~~~~  if     $(q', b,S) \in \delta(q, a)$                                            \\
qa  & bq'       ~~~~ ~~~~~~  if     $(q', b,R) \in \delta(q, a)$             \\
cqa & q'cb        ~~~ ~~~~~~  if     $(q', b,L) \in \delta(q, a)$                                                        \\
q\# & bq'\#      ~~ ~~~~~~  if     $(q', b, R) \in \delta(q, \Box)$                                                                     \\
cq\#  & q'cb\#           ~ ~~~~~~  if     $(q', b, L) \in \delta(q, \Box)$                                                                   \\
q\#  &  q'b\#                 ~~ ~~~~~~  if     $(q', b, S) \in \delta(q, \Box)$                                                              \\

\end{tabular}

\hs The integer $n$ is the number of words in the list $x$ and also in the list $y$. We assume that these two lists are indexed so that 
$x=(x_i)_{1\leq i \leq n}$ and $y=(y_i)_{1\leq i \leq n}$.  Let now $E \subseteq \{1, 2, \ldots , n\}$ be the set of integers $i$ such that 
the initial state $q_0$ of the Turing machine $\mathcal{M}$ appears in the word $y_i$. The $\om$-language $L \subseteq \{1, 2, \ldots , n\}^\om$ 
of infinite words over the alphabet $\{1, 2, \ldots , n\}$ which begin by the letter $1$ and have infinitely many letters in $E$ 
is  a regular $\om$-language and it is accepted by a 
(deterministic) B\"uchi automaton $\mathcal{A}$. We now consider the instance $I=((x_1,\ldots ,x_n), (y_1,\ldots ,y_n), \mathcal{A})$ 
of the $\om$-PCP(Reg). It is easy to check  that  this instance of the $\om$-PCP(Reg) has a solution  $i_1, i_2,\ldots ,i_k \ldots $ if and only if 
the Turing machine $\mathcal{M}$, when started on a blank tape, admits an infinite 
computation that reenters infinitely often in the initial state $q_0$. 

\hs Thus the $\Sigma_1^1$-complete problem $(P)$ is  reduced to the $\om$-PCP in a regular $\om$-language and this latter problem is also 
$\Sigma_1^1$-complete. 
\ep

\section{Applications to infinitary rational relations}

\subsection{Infinitary rational relations}

\noi We now  recall the definition  of infinitary rational relations,  via definition by B\"uchi transducers:

\begin{Deff}
  A B\"uchi transducer is a sextuple $\mathcal{T}=(K, \Si, \Ga, \Delta,
  q_0, F)$, where $K$ is a finite set of states, $\Si$ and $\Ga$ are finite
  sets called the input and the output alphabets, $\Delta$ is a finite
  subset of $K \times \Sis \times \Gas \times K$ called the set of
  transitions, $q_0$ is the initial state, and $F \subseteq K$ is the set
  of accepting states.  \nl A computation $\mathcal{C}$ of the transducer
  $\mathcal{T}$ is an infinite sequence of consecutive transitions
  \begin{displaymath}
  (q_0, u_1, v_1, q_1), (q_1, u_2, v_2, q_2), \ldots 
  (q_{i-1}, u_{i}, v_{i}, q_{i}),  (q_i, u_{i+1}, v_{i+1}, q_{i+1}), \ldots 
  \end{displaymath}
  The computation is said to be successful iff there exists a final state
  $q_f \in F$ and infinitely many integers $i\geq 0$ such that $q_i=q_f$.
  The input word and output word of the computation are respectively
  $u=u_1.u_2.u_3 \ldots$ and $v=v_1.v_2.v_3 \ldots$ The input and the
  output words may be finite or infinite.  The infinitary rational relation
  $R(\mathcal{T})\subseteq \Sio \times \Ga^\om$ accepted by the B\"uchi
  transducer $\mathcal{T}$ is the set of couples $(u, v) \in \Sio \times
  \Ga^\om$ such that $u$ and $v$ are the input and the output words of some
  successful computation $\mathcal{C}$ of $\mathcal{T}$.  The set of
  infinitary rational relations will be denoted $RAT_2$.
\end{Deff} 

\noi If $R(\mathcal{T})\subseteq \Sio \times \Ga^\om$ is an infinitary rational
relation recognized by the B\"uchi transducer $\mathcal{T}$ then we denote
$$Dom(R(\mathcal{T}))=\{ u \in \Sio \mid \exists v \in \Gao ~~(u, v) \in
R(\mathcal{T}) \}$$
\noi
 and 
$$Im(R(\mathcal{T}))=\{ v \in \Gao \mid \exists u \in
\Sio (u, v) \in R(\mathcal{T}) \}.$$

\noi It is well known that, for each infinitary rational relation
$R(\mathcal{T})\subseteq \Sio \times \Ga^\om$, the sets
$Dom(R(\mathcal{T}))$ and $Im(R(\mathcal{T}))$ are regular $\om$-languages and that one can construct, from the 
B\"uchi transducer $\mathcal{T}$, some B\"uchi automata $\mathcal{A}$ and $\mathcal{B}$ accepting the $\om$-languages 
$Dom(R(\mathcal{T}))$ and $Im(R(\mathcal{T}))$. 

\hs To each B\"uchi transducer  $\mathcal{T}$ can be associated in an injective and recursive way an index $z\in \mathbb{N}$ and we shall denote 
in the sequel  $\mathcal{T}_z$ the  B\"uchi transducer of index $z$. 

\hs We proved in \cite{Fin-HI} that many decision problems about infinitary rational relations are highly undecidable. In fact many of them, like 
the universality problem, the equivalence problem, the inclusion problem, the cofiniteness problem, the unambiguity problem,  
are $\Pi_2^1$-complete, hence located at the second level of the analytical hierarchy. 

\hs We can now use the $\Sigma_1^1$-completeness of the $\om$-PCP in a regular $\om$-language to obtain a new result of high undecidability.

\begin{The}
It is $\Pi_1^1$-complete to determine whether two given infinitary rational relations are disjoint, i.e. the set 
$\{ (z, z') \in \mathbb{N}^2  \mid R(\mathcal{T}_z) \cap R(\mathcal{T}_{z'}) = \emptyset \}$ is $\Pi_1^1$-complete. 
\end{The}

\proo  We are going to show that the complement of this set is $\Sigma_1^1$-complete, i.e. that the set 
$\{ (z, z') \in \mathbb{N}^2  \mid R(\mathcal{T}_z) \cap R(\mathcal{T}_{z'}) \neq \emptyset \}$ is $\Sigma_1^1$-complete. 

\hs Firstly, it is easy to see that, for two given B\"uchi transducers $\mathcal{T}_z$ and $\mathcal{T}_{z'}$, one can define a B\"uchi Turing machine 
$\mathcal{M}_{\Phi(z, z')}$ of index $\Phi(z, z')$ accepting the $\om$-language $R(\mathcal{T}_z) \cap R(\mathcal{T}_{z'})$. 
Moreover one can construct the function $\Phi: \mathbb{N}^2 \ra \mathbb{N}$ as an injective computable function. This shows that the set 
$\{ (z, z') \in \mathbb{N}^2  \mid R(\mathcal{T}_z) \cap R(\mathcal{T}_{z'}) \neq \emptyset \}$ is reduced to the set 
$\{ z \in \mathbb{N} \mid L(\mathcal{M}_z) \neq \emptyset \}$ which is in the class $\Sigma_1^1$, since the non-emptiness problem for $\om$-languages 
of Turing machines is in the class $\Sigma_1^1$. Thus the set $\{ (z, z') \in \mathbb{N}^2  \mid R(\mathcal{T}_z) \cap R(\mathcal{T}_{z'}) \neq \emptyset \}$
is in the class $\Sigma_1^1$. 

\hs Secondly, we have to show the completeness part of the theorem. We are going to reduce the $\om$-PCP in a regular $\om$-language 
to the problem of the  non-emptiness of the intersection of two infinitary rational relations. Let then 
 $I=((x_1,\ldots ,x_n), (y_1,\ldots ,y_n), \mathcal{A})$ be an instance of the $\om$-PCP(Reg), where the $x_i$ and $y_i$ are words over an 
alphabet $\Ga$. We can then construct B\"uchi transducers $\mathcal{T}_{\psi_1(\bar{I})}$ and $\mathcal{T}_{\psi_2(\bar{I})}$ such that 
the  infinitary rational relation $R(\mathcal{T}_{\psi_1(\bar{I})}) \subseteq \{1, 2, \ldots, n\}^\om \times \Gao$ is the set of  pairs of infinite words in the form 
$(i_1 i_2 i_3 \cdots ; x_{i_1}x_{i_2}x_{i_3} \cdots )$ with $i_1 i_2 i_3 \cdots \in  L(\mathcal{A})$. And similarly 
$R(\mathcal{T}_{\psi_2(\bar{I})}) \subseteq \{1, 2, \ldots, n\}^\om \times \Gao$ is the set of  pairs of infinite words in the form 
$(i_1 i_2 i_3 \cdots ; y_{i_1}y_{i_2}y_{i_3} \cdots )$ with $i_1 i_2 i_3 \cdots \in  L(\mathcal{A})$. 
Thus it holds that $R(\mathcal{T}_{\psi_1(\bar{I})}) \cap R(\mathcal{T}_{\psi_2(\bar{I})})$ is non-empty iff there is an infinite sequence 
$i_1i_2 \cdots i_k \cdots \in L(\mathcal{A})$ such that 
$x_{i_1}x_{i_2}\cdots x_{i_k}\cdots  =y_{i_1}y_{i_2}\cdots y_{i_k}\cdots $. 
The reduction is now given by the injective computable function $\Psi: \mathbb{N} \ra \mathbb{N}^2$ given by $\Psi(z)=(\Psi_1(z), \Psi_2(z))$. 
\ep

\subsection{Continuity  of   $\om$-rational functions}

\hs 
Recall that an infinitary rational relation $R(\mathcal{T})\subseteq \Sio \times \Ga^\om$ is said to be functional iff it is the graph of a function,  i.e. iff 
$$[\fa x \in Dom(R(\mathcal{T})) ~~ \exists ! y \in Im(R(\mathcal{T})) ~~ (x, y) \in R(\mathcal{T})].$$ 

\noi Then the  functional relation $R(\mathcal{T})$  defines an $\om$-rational (partial) function
$F_{\mathcal{T}}: Dom(R(\mathcal{T})) \subseteq \Sio \ra \Gao$ by:
 for each $u\in Dom(R(\mathcal{T}))$, $F_{\mathcal{T}}(u)$ is the unique
$v\in\Gao$ such that $(u, v) \in R(\mathcal{T})$.

\hs Recall the following previous decidability result. 

\begin{The}[\cite{Gire86}]
  One can decide whether an infinitary rational relation recognized by a
  given B\"uchi transducer $\mathcal{T}$ is a functional infinitary
  rational relation.
\end{The}

\noi One can then associate in a recursive and injective way an index  to each  B\"uchi transducer $\mathcal{T}$ accepting a 
functional infinitary rational relation $R(\mathcal{T})$. In the sequel we consider only these B\"uchi transducers and we shall denote 
$\mathcal{T}_z$ the B\"uchi transducer of index $z$ (such that $R(\mathcal{T}_z)$ is functional). 

\hs  It is  very natural to consider the notion of continuity for  $\om$-rational  functions defined by B\"uchi transducers. 

\hs We assume the reader to be familiar with basic notions of topology which
may be found in \cite{Kechris94,Thomas90,Staiger97,PerrinPin}.  There is a natural metric on
the set $\Sio$ of infinite words over a finite alphabet $\Si$ which is
called the prefix metric and defined as follows. For $u, v \in \Sio$ and
$u\neq v$ let $d(u, v)=2^{-l_{pref(u,v)}}$ where $l_{pref(u,v)}$ is the
least integer $n$ such that the $(n+1)^{th}$ letter of $u$ is different
from the $(n+1)^{th}$ letter of $v$.  This metric induces on $\Sio$ the
usual Cantor topology for which open subsets of $\Sio$ are in the form
$W\cdot \Si^\om$, where $W\subseteq \Sis$.

\hs We recall that a function $f: Dom(f) \subseteq
\Sio \ra \Gao$, whose domain is $Dom(f)$, is said to be continuous at point
$x\in Dom(f)$ if :

$$\forall n\geq 1 ~~~\exists k \geq 1 ~~~\forall y\in Dom(f) ~~~[~ d(x, y)<2^{-k} \Rightarrow d(f(x), f(y))<2^{-n}~] $$

\noi  The continuity set $C(f)$ of the function $f$ is the set of points of
continuity of $f$. The function $f$ is said to be continuous if it is continuous at every
point $x\in Dom(f)$, i. e. if $C(f)=Dom(f)$.

\hs Prieur proved the following decidability result. 

\begin{The}[Prieur \cite{Prieur01, Prieur02}]
   One can decide whether a given $\om$-rational function is continuous. 
\end{The}

\noi On the other hand   the following undecidability result was proved in  \cite{CFS08}. 

\begin{The}[see  \cite{CFS08}]
  One cannot decide whether a given
  $\om$-rational function $f$ has at least one point of continuity.
\end{The}

\noi We can now give the exact complexity of this undecidable problem. 

\begin{The}\label{mainthe}
 It is  $\Sigma_1^1$-complete   to    determine  whether a given    
$\om$-rational function $f$ has at least one point of continuity, i.e. whether 
the continuity set $C(f)$ of $f$ is non-empty.  In other words the set 
$\{ z\in \mathbb{N} \mid  C(F_{\mathcal{T}_z}) \neq \emptyset  \}$ is $\Sigma_1^1$-complete. 
\end{The}

\noi We first prove the following lemma. 

\begin{Lem}\label{lem1}
The set $\{ z\in \mathbb{N} \mid  C(F_{\mathcal{T}_z}) \neq \emptyset  \}$ is in the class $\Sigma_1^1$. 
\end{Lem}

\proo 
 Let $F$ be a  function from $Dom(F) \subseteq \Sio $ into $\Gao$. For some integers $n, k\geq 1$, we consider 
the set 
$$X_{k,n}=\{ x\in Dom(F) \mid 
\forall y\in Dom(F) ~~~[~ d(x, y)<2^{-k} \Rightarrow d(F(x), F(y))<2^{-n}~] \}$$

\noi  For $x\in Dom(F)$ it holds that: 
$$ x \in C(F) \Longleftrightarrow \forall n\geq 1 ~~~\exists k \geq 1 ~~~[~ x \in  X_{k,n}~] $$

\noi We shall denote $X_{k,n}(z)$  
the set $X_{k,n}$ corresponding to the function $F_{\mathcal{T}_z}$ defined by the B\"uchi transducer of index $z$. Then 
 it holds that: 
$$ x \in C(F_{\mathcal{T}_z}) \Longleftrightarrow \forall n\geq 1 ~~~\exists k \geq 1 ~~~[~ x \in  X_{k,n}(z)~] $$

\noi And we denote $R_{k, n}(x, z) $ the relation given by: 
$$R_{k, n}(x, z)  \Longleftrightarrow [~ x \in  X_{k,n}(z)~] $$

\noi We now prove that this relation is  a $\Pi_3^0$-relation.  

\hs For $x\in\Sio$ and $k\in \mathbb{N}$, 
we denote $B(x, 2^{-k})$  the open ball of center $x$ and 
of radius $2^{-k}$, i.e. the set of $y\in \Sio$ such that $d(x, y)<2^{-k}$. 
We know, from  the definition of the distance $d$, 
that for two $\om$-words $x$ and  $y$ over $\Si$, the inequality $d(x, y)<2^{-k}$ simply means that $x$ and  $y$ 
have the same $(k+1)$ first letters. Thus $B(x, 2^{-k})=x[k+1]\cdot\Sio$. But by definition of $X_{k,n}(z)$ it holds that: 

\hs $x \in  X_{k,n}(z)$
$\nl  \Longleftrightarrow ( x\in  Dom(F_{\mathcal{T}_z}) \mbox{ and } 
F_{\mathcal{T}_z}[ B(x, 2^{-k})\cap  Dom(F_{\mathcal{T}_z}) ] \subseteq B( F_{\mathcal{T}_z}(x), 2^{-n} )  )$

\hs We claim that there is an algorithm which, given $x\in \Sio$ and $z\in \mathbb{N}$, 
can decide whether 
$$F_{\mathcal{T}_z}[ B(x, 2^{-k})\cap  Dom(F_{\mathcal{T}_z}) ] \subseteq w\cdot \Gao,$$
\noi for some finite word $w\in\Gas$ such that $|w|=n+1$. 

\hs  Indeed the $\om$-language 
$B(x, 2^{-k})\cap  Dom(F_{\mathcal{T}_z}) = x[k+1]\cdot\Sio \cap  Dom(F_{\mathcal{T}_z})$ is the intersection of two regular $\om$-languages and 
one can construct a B\"uchi  automaton accepting it. The graph of the restriction of the function $F_{\mathcal{T}_z}$ to the set 
$x[k+1]\cdot\Sio \cap  Dom(F_{\mathcal{T}_z})$ is also an infinitary rational relation and one can then also find a 
B\"uchi  automaton $\mathcal{B}$ accepting $F_{\mathcal{T}_z}[ x[k+1]\cdot\Sio \cap  Dom(F_{\mathcal{T}_z})  ] $. 
One can then find the set 
of prefixes of length $n+1$ of infinite words in $L(\mathcal{B})$. If there is only one such prefix $w$ then 
$F_{\mathcal{T}_z}[ B(x, 2^{-k})\cap  Dom(F_{\mathcal{T}_z}) ] \subseteq w\cdot \Gao$ and otherwise we have 
$F_{\mathcal{T}_z}[ B(x, 2^{-k})\cap  Dom(F_{\mathcal{T}_z}) ] \nsubseteq w'\cdot \Gao$ for every word $w'\in\Gas$ such that $|w'|=n+1$. 
We now write $S(x, k, n, z)$ iff $F_{\mathcal{T}_z}[ B(x, 2^{-k})\cap  Dom(F_{\mathcal{T}_z}) ] \subseteq w\cdot \Gao,$
 for some finite word $w\in\Gas$ such that $|w|=n+1$. As we have just seen the relation $S(x, k, n, z)$ is computable, i.e. a $\Delta_1^0$ relation. 

\hs  On the other hand, we have 
$$x \in  X_{k,n}(z)  \Longleftrightarrow ( x\in  Dom(F_{\mathcal{T}_z}) \mbox{ and } S(x, k, n, z) )$$

\noi But $Dom(F_{\mathcal{T}_z})$ is a regular $\om$-language accepted by a B\"uchi automaton $\mathcal{A}$ which can be constructed effectively 
from $\mathcal{T}_z$ and hence from the index $z$. And the relation $( x\in L(\mathcal{A}) )$ is known to be an arithmetical 
 $\Pi_3^0$ (and also a $\Sigma_3^0$) relation, see \cite{LescowThomas}. Thus ``$x \in  X_{k,n}(z) $" can be expressed also by a  
$\Pi_3^0$ (and also a $\Sigma_3^0$) formula because the relation $S$ is a $\Delta_1^0$ relation. 

\hs Now we have the following equivalences: 
$$C(F_{\mathcal{T}_z})\neq \emptyset  \Longleftrightarrow \exists x ~~[ x  \in C(F_{\mathcal{T}_z}) ] \Longleftrightarrow 
\exists x  ~~[  \forall n \geq 1 ~~\exists k \geq 1 ~~x \in  X_{k,n}(z)   ] $$

\noi Clearly the formula $\exists x  ~~[  \forall n \geq 1 ~~\exists k \geq 1 ~~ x \in  X_{k,n}(z)   ] $ is a $\Sigma_1^1$-formula where there is a second order 
quantification $\exists x$ followed by an arithmetical  $\Pi_5^0$-formula in which  the quantifications $ \forall n \geq 1 ~~\exists k \geq 1$ 
are first order quantifications on integers.  
\ep 

\hs {\bf End of Proof of Theorem \ref{mainthe}.}  To prove the completeness part of the theorem we  use 
some ideas of \cite{CFS08} but we shall modify 
the constructions of  \cite{CFS08} in order to use the $\Sigma_1^1$-completeness of the $\om$-PCP(Reg) instead of the undecidability of the PCP. 
We are  now going to  reduce the $\om$-PCP in a regular $\om$-language 
to the non-emptiness of the continuity set of an $\om$-rational function.  

\hs Let then  
 $I=((x_1,\ldots ,x_n), (y_1,\ldots ,y_n), \mathcal{A})$ be an instance of the $\om$-PCP(Reg), where the $x_i$ and $y_i$ are words over an 
alphabet $\Ga$. We can  construct an $\om$-rational function $F$   in the following way. 

\hs Firstly,  the domain $Dom(F)$ will be a 
 set of $\om$-words over the alphabet $\{ 1, \ldots, n \} \cup \{ a, b \}$, where $a$ and $b$ 
are new letters not in $\{ 1, \ldots, n \}$. For $x \in ( \{ 1, \ldots, n \} \cup \{  a, b \} )^\om$  
we denote $x( / \{ a, b \} )$  the (finite or infinite) word over the alphabet  $\{ 1, \ldots, n \}$ obtained from $x$ 
when removing every occurrence of the letters $a$ and $b$. And  $x( / \{1, \ldots, n \} )$ is the (finite or infinite) word over the alphabet  $\{ a, b \}$ 
obtained from $x$ when removing every occurrence of the letters $1, \ldots, n$.
Then $Dom(F)$ is the set of 
$\om$-words $x$ over the alphabet $\{ 1, \ldots, n \} \cup \{ a, b \}$ such that $x( / \{ a, b \} ) \in L(\mathcal{A})$ (so in particular $x( / \{ a, b \} )$ is infinite) 
{\bf and} $x( / \{1, \ldots, n \} )$ is infinite.  
It is clear that this domain is a regular $\om$-language. 

\hs Secondly, for $x\in Dom(F)$ such that $x( / \{ a, b \} )=i_1i_2 \cdots i_k \cdots \in L(\mathcal{A})$ we set:  
\begin{itemize}
  \item
$F(x)=x_{i_1}x_{i_2}\cdots x_{i_k}\cdots$~~~~~~~~ if $x( / \{1, \ldots, n \} ) \in (\{ a, b \}^\star\cdot a)^\om$, and 

  \item $F(x)=y_{i_1}y_{i_2}\cdots y_{i_k}\cdots$ ~~~~~~~~ if $x( / \{1, \ldots, n \} ) \in \{ a, b \}^\star\cdot b^\om$. 
  \end{itemize}

  \noi The $\om$-language  $(\{ a, b \}^\star.a)^\om$ is  the set of $\om$-words over the alphabet $\{ a, b \}$ having infinitely many  
 letters $a$. The $\om$-language  $\{ a, b \}^\star.b^\om$ is the complement in   $\{ a, b \}^\om$    of the $\om$-language  $ (\{ a, b \}^\star.a)^\om$: 
it is the set  of $\om$-words over the alphabet $\{ a, b \}$ containing only 
  finitely many  letters $a$. 
The two $\om$-languages $(\{ a, b \}^\star.a)^\om$ and $\{ a, b \}^\star.b^\om$ are
  $\om$-regular, and one can easily construct  B\"uchi automata accepting them.  
Then it is easy
  to see that the function $F$ is $\om$-rational and that one can construct a
 B\"uchi transducer $\mathcal{T}$ accepting  the graph of the function $F$. Moreover one can construct an injective computable function 
$\psi: \mathbb{N} \ra  \mathbb{N}$ such that $\mathcal{T}=\mathcal{T}_{\psi(\bar{I})}$ and so $F=F_{\mathcal{T}_{\psi(\bar{I})}}$.

 \hs We now prove that if $x \in Dom(F_{\mathcal{T}_{\psi(\bar{I})}})$  is a point of continuity of the function $F_{\mathcal{T}_{\psi(\bar{I})}}$ 
then the $\om$-PCP(Reg)  of instance 
 $I=((x_1,\ldots ,x_n), (y_1,\ldots ,y_n), \mathcal{A})$ has  a solution $i_1i_2 \cdots i_k \cdots$, i.e.  an $\om$-word 
$i_1i_2 \cdots i_k \cdots \in L(\mathcal{A})$ such that 
$x_{i_1}x_{i_2}\cdots x_{i_k}\cdots  =y_{i_1}y_{i_2}\cdots y_{i_k}\cdots $.

\hs  To simplify the notations we denote by $F$ the function $F_{\mathcal{T}_{\psi(\bar{I})}}$. 
We now distinguish two cases. 

  \paragraph{First Case.} Assume firstly that $x( / \{1, \ldots, n \} ) \in (\{ a, b \}^\star\cdot a)^\om$ and that 
$x( / \{ a, b \} )=i_1i_2 \cdots i_k \cdots \in L(\mathcal{A})$. 
  Then by definition of $F$ it holds that $F(x)=x_{i_1}x_{i_2}\cdots x_{i_k}\cdots$. We denote $z = x( / \{1, \ldots, n \} )$. 
  Notice that there is a sequence of elements $z_p \in  \{ a, b \}^\star\cdot b^\om$, $p\geq 1$, such that  the sequence $(z_p)_{p\geq 1}$ is convergent and 
  $lim (z_p) = z = x( / \{1, \ldots, n \} )$. This is due to the fact that $\{ a, b \}^\star.b^\om$ is dense in  $\{ a, b \}^\om$. 
  We call $t_p$ the infinite word over the alphabet $\{ 1, \ldots, n \} \cup \{ a, b \}$  such that,  for each integer $i\geq 1$,  we have 
$t_p(i)=x(i)$ if $x(i) \in \{ 1, \ldots, n \}$ and $t_p(i)=z_p(k)$  if $x(i)$ is the $k$ th letter of $z$. Then the sequence $(t_p)_{p\geq 1}$ is convergent and 
  $lim (t_p) = x$. But by definition of $F$ it holds that $F(t_p)=y_{i_1}y_{i_2}\cdots y_{i_k}\cdots$   for every integer $p\geq 1$ while 
  $F(x)=x_{i_1}x_{i_2}\cdots x_{i_k}\cdots$. Thus if $x$ is a point of continuity of the function $F$ then it holds that 
$x_{i_1}x_{i_2}\cdots x_{i_k}\cdots = y_{i_1}y_{i_2}\cdots y_{i_k}\cdots$ and the $\om$-PCP(Reg)  of instance 
 $I=((x_1,\ldots ,x_n), (y_1,\ldots ,y_n), \mathcal{A})$ has  a solution $i_1i_2 \cdots i_k \cdots$. 

\paragraph{Second  Case.} 
Assume now that $x( / \{1, \ldots, n \} ) \in \{ a, b \}^\star\cdot b^\om$ and that 
$x( / \{ a, b \} )=i_1i_2 \cdots i_k \cdots \in L(\mathcal{A})$. 
 Notice that $(\{ a, b \}^\star.a)^\om$ is also dense in $\{ a, b \}^\om$. Then reasoning as in the first case 
    we can prove that if $x$ is a
   point of continuity of $F$ then $x_{i_1}x_{i_2}\cdots x_{i_k}\cdots = y_{i_1}y_{i_2}\cdots y_{i_k}\cdots$ and the $\om$-PCP(Reg)  of instance 
 $I=((x_1,\ldots ,x_n), (y_1,\ldots ,y_n), \mathcal{A})$ has  a solution $i_1i_2 \cdots i_k \cdots$. 

\hs    Conversely assume that the  $\om$-PCP in a regular $\om$-language   of instance 
 $I=((x_1,\ldots ,x_n), (y_1,\ldots ,y_n), \mathcal{A})$ has  a solution $i_1i_2 \cdots i_k \cdots$, i.e.  an $\om$-word 
$i_1i_2 \cdots i_k \cdots \in L(\mathcal{A})$ such that 
$x_{i_1}x_{i_2}\cdots x_{i_k}\cdots  =y_{i_1}y_{i_2}\cdots y_{i_k}\cdots $.   
We now show that each $x\in Dom(F)$ such that $x( / \{ a, b \} )=i_1i_2 \cdots i_k \cdots$ is a point of continuity of 
the function $F$. Consider an infinite sequence $(t_p)_{p\geq 1}$ of elements of $Dom(F)$ such that $lim(t_p)=x$. It is easy to see 
that the sequence $(t_p( / \{ a, b \} ))_{p\geq 1}$ is convergent and that its limit is the $\om$-word $x( / \{ a, b \} )=i_1i_2 \cdots i_k \cdots$. 
This implies easily that 
the sequence $F(t_p)_{p\geq 1}$ is convergent and that its limit is the $\om$-word 
$F(x)=x_{i_1}x_{i_2}\cdots x_{i_k}\cdots  = y_{i_1}y_{i_2}\cdots y_{i_k}\cdots $. 
Thus $x$ is a point of continuity of $F$ and this ends the proof. 
\ep

\hs We consider now the continuity set of an $\om$-rational function and its possible complexity. 
The following undecidability result was proved in  \cite{CFS08}. 

\begin{The}[see  \cite{CFS08}]\label{reg}
  One cannot decide whether the continuity set of a  given
  $\om$-rational function $f$ is a regular (respectively, context-free) $\om$-language.  
\end{The}

\noi We can now give the exact complexity of the first above  undecidable problem. 

\begin{The}\label{mainthe2}
 It is  $\Pi_1^1$-complete   to    determine  whether the continuity set $C(f)$ of a given    
$\om$-rational function $f$ is a regular $\om$-language.  In other words the set 
$\{ z\in \mathbb{N} \mid  C(F_{\mathcal{T}_z}) \mbox{ is a regular }  \om \mbox{-language} \}$ is $\Pi_1^1$-complete. 
\end{The}

\noi We first prove the following lemma. 

\begin{Lem}\label{regP}
The set $\{ z\in \mathbb{N} \mid  C(F_{\mathcal{T}_z})  \mbox{ is a regular }  \om \mbox{-language} \}$ is in the class $\Pi_1^1$. 
\end{Lem}

\proo
Recall that $\mathcal{A}_z$ denotes the B\"uchi automaton of index $z$. We can  express the sentence 
``$C(F_{\mathcal{T}_z})$  is a regular   $\om$-language" by the sentence: 
$$\exists z' ~~ C(F_{\mathcal{T}_z}) = L(\mathcal{A}_{z'}).$$
\noi On the other hand we have seen in the proof of Lemma \ref{lem1} that 
 $$x \in C(F_{\mathcal{T}_z})  \Longleftrightarrow [  \forall n \geq 1 ~~\exists k \geq 1 ~~x \in  X_{k,n}(z)   ] $$
\noi and then that  $x \in C(F_{\mathcal{T}_z}) $ can be expressed  
 by an arithmetical  $\Pi_5^0$-formula. 
We can now express $C(F_{\mathcal{T}_z}) = L(\mathcal{A}_{z'})$ by:
$$\fa x ~~ [ ( x \in C(F_{\mathcal{T}_z}) \mbox{ and } x \in L(\mathcal{A}_{z'}) ) \mbox{ or  } 
( x \notin C(F_{\mathcal{T}_z}) \mbox{ and } x \notin L(\mathcal{A}_{z'}) ) ]$$ 
\noi which is a $\Pi_1^1$-formula because there is one universal second order quantification $\fa x$ followed by an arithmetical formula 
(recall that $x \in L(\mathcal{A}_{z'}) $ can be expressed by an  arithmetical $\Pi_3^0$-formula). 

\hs Finally the sentence $$\exists z' ~~ C(F_{\mathcal{T}_z}) = L(\mathcal{A}_{z'})$$
\noi can be expressed by a $\Pi_1^1$-formula because the quantification $\exists z' $ is a first-order quantification bearing on integers 
and the formula $C(F_{\mathcal{T}_z}) = L(\mathcal{A}_{z'})$ can be expressed by a  $\Pi_1^1$-formula. 
\ep

\hs {\bf End of Proof of Theorem \ref{mainthe2}.}  To prove the completeness part of the theorem we reduce the $\om$-PCP in a regular $\om$-language 
to the problem of the non-regularity  of the continuity set of an $\om$-rational function. 

\hs As in the proof of the above Theorem \ref{reg} in \cite{CFS08}, we shall use a particular instance of Post Correspondence Problem. For
  two letters $c, d$, let PCP$_1$ be the Post Correspondence Problem of
  instance $((t_1, t_2, t_3),(w_1, w_2, w_3))$, where $t_1=c^2$,
  $t_2=t_3=d$ and $w_1=w_2=c$, $w_3=d^2$. It is easy to see that its
  solutions are the sequences of indices in $\{ 1^{i}\cdot 2^{i}\cdot 3^{i} \mid
  i\geq 1 \} \cup \{ 3^{i}\cdot 2^{i}\cdot 1^{i} \mid i\geq 1 \}$.  In particular, 
  this language over the alphabet $\{1, 2, 3\}$ is not context-free and
  this will be useful in the sequel. 

\hs Let then  
 $I=((x_1,\ldots ,x_n), (y_1,\ldots ,y_n), \mathcal{A})$ be an instance of the $\om$-PCP(Reg), where the $x_i$ and $y_i$ are words over an 
alphabet $\Ga$. We can  construct an $\om$-rational function $F'$   in the following way. 

\hs Let $D=\{d_1, d_2 , d_3\}$ such that $D$ and $\{ 1, \ldots, n \} \cup \{ a, b \}$ are disjoint. 
The domain $Dom(F')$ will be a 
 set of $\om$-words in $D^+\cdot Dom(F)$, where, as in the proof of Theorem \ref{mainthe}, $Dom(F)$ is
 the set of 
$\om$-words $x$ over the alphabet $\{ 1, \ldots, n \} \cup \{ a, b \}$ such that $x( / \{ a, b \} ) \in L(\mathcal{A})$ (so in particular $x( / \{ a, b \} )$ is infinite) 
{\bf and} $x( / \{1, \ldots, n \} )$ is infinite.  
It is clear that the domain $Dom(F')$  is a regular $\om$-language.

\hs Now, for $x\in Dom(F')$ such that $x=d_{j_1}\cdots d_{j_p}\cdot y$ with $y\in Dom(F)$ and 
$y( / \{ a, b \} )=i_1i_2 \cdots i_k \cdots \in L(\mathcal{A})$ we set: 
 
\begin{itemize}
  \item
$F'(x)=t_{j_1}\cdots t_{j_p}x_{i_1}x_{i_2}\cdots x_{i_k}\cdots$~~~~~~~~ if $y( / \{1, \ldots, n \} ) \in (\{ a, b \}^\star\cdot a)^\om$, and 

  \item $F'(x)=w_{j_1}\cdots w_{j_p}y_{i_1}y_{i_2}\cdots y_{i_k}\cdots$ ~~~~~~~~ if 
$y( / \{1, \ldots, n \} ) \in \{ a, b \}^\star\cdot b^\om$. 
  \end{itemize}

\noi Then it is easy
  to see that the function $F'$ is $\om$-rational and that one can construct a
 B\"uchi transducer $\mathcal{T'}$ accepting  the graph of the function $F'$. Moreover one can construct an injective computable function 
$\Theta: \mathbb{N} \ra  \mathbb{N}$ such that $\mathcal{T'}=\mathcal{T}_{\Theta(\bar{I})}$ and so $F'=F_{\mathcal{T}_{\Theta(\bar{I})}}$.

\hs Reasoning as in the preceding proof 
we can
  prove that the function $F'$ is continuous at point 
$x=d_{j_1}\cdots d_{j_p}\cdot y$, where $y\in Dom(F)$ , if and only if the 
 the  sequence $j_1, \ldots , j_p$ is a solution of the Post Correspondence
  Problem PCP$_1$ and    $y( / \{ a, b \} )=i_1i_2 \cdots i_k \cdots $  is a solution of 
the $\om$-PCP(Reg)  of instance 
 $I=((x_1,\ldots ,x_n), (y_1,\ldots ,y_n), \mathcal{A})$.

\hs Thus  if the $\om$-PCP(Reg)  of instance 
 $I=((x_1,\ldots ,x_n), (y_1,\ldots ,y_n), \mathcal{A})$ has no solution, then the continuity set 
$C(F')$ is empty, hence it is $\om$-regular. 

\hs On the other hand assume that the $\om$-PCP(Reg)  of instance 
 $I=((x_1,\ldots ,x_n)$, $(y_1,\ldots ,y_n), \mathcal{A})$ has some solutions. In that case the continuity set 
$C(F')$ is in the form $T\cdot R$ where $T = \{ d_1^{i}\cdot d_2^{i}\cdot d_3^{i} \mid
  i\geq 1 \} \cup \{ d_3^{i}\cdot d_2^{i}\cdot d_1^{i} \mid i\geq 1 \}$ and $R$ is a set of infinite words over the alphabet 
$\{ 1, \ldots, n \} \cup \{ a, b \}$. In that case the continuity set $C(F')$ can not be $\om$-regular because otherwise the language 
$T$ should be regular (since $D=\{d_1, d_2 , d_3\}$ and $\{ 1, \ldots, n \} \cup \{ a, b \}$ are disjoint) and it is not even context-free.

\hs This shows that the $\om$-PCP(Reg)  of instance 
 $I=((x_1,\ldots ,x_n), (y_1,\ldots ,y_n), \mathcal{A})$ has a solution if and only if  the continuity set 
$C(F')$ is not $\om$-regular. This ends the proof. 
\ep

\hs It is natural to ask whether the set $\{ z\in \mathbb{N} \mid  C(F_{\mathcal{T}_z})  \mbox{ is a context-free }$  $\om$-language$\}$ is also 
$\Pi_1^1$-complete.  But one cannot extend directly Lemma \ref{regP},  replacing regular by context-free.  If we replace the B\"uchi automaton 
$\mathcal{A}_z$  of index $z$ by the B\"uchi pushdown automaton    $\mathcal{B}_z$  of index $z$,   we get only that the set 
$\{ z\in \mathbb{N} \mid  C(F_{\mathcal{T}_z})  \mbox{ is a context-free }  \om \mbox{-language} \}$ is in the class $\Pi_2^1$ because the 
``$x \in L(\mathcal{B}_{z'}) $" can only be expressed by a $\Sigma_1^1$-formula. On the other hand, the second part of 
the proof of Theorem \ref{mainthe2} proves in the same way that 
the set $\{ z\in \mathbb{N} \mid  C(F_{\mathcal{T}_z})  \mbox{ is a context-free }  \om \mbox{-language} \}$  is $\Pi_1^1$-hard. Thus we can now state 
the following result. 

\begin{The}\label{mainthe3}
The set 
$\{ z\in \mathbb{N} \mid  C(F_{\mathcal{T}_z}) \mbox{ is a context-free }  \om \mbox{-language} \}$ is $\Pi_1^1$-hard 
and in the class $\Pi_2^1 \setminus \Sigma_1^1$. 
\end{The}

\section{Concluding remarks}

\noi We have given a complete proof of the $\Sigma_1^1$-completeness of 
 the $\om$-PCP in a regular $\om$-language, also denoted $\om$-PCP(Reg).  Then we have applied this result and obtained the exact complexity of several 
highly undecidable problems about infinitary rational relations and $\om$-rational functions. In particular, we have showed that 
there is an amazing gap between two decision problems about $\om$-rational functions realized by 
finite state B\"uchi transducers:   
 it is decidable whether a given $\om$-rational function is continuous, while  it is 
 $\Sigma_1^1$-complete to determine whether a given $\om$-rational function has at least one point of continuity. 

\hs We hope that this paper will attract the reader's attention on a new highly undecidable problem, the  $\om$-PCP(Reg), which could be very useful to study 
the frontiers between decidable and undecidable problems.

\end{document}